\documentclass[12pt,twoside]{article}
\usepackage{amssymb}
\usepackage{amsmath}
\usepackage{latexsym}
\usepackage{longtable}
\usepackage{epsfig}
\usepackage{graphicx,bbm,psfrag}
\graphicspath{{images/}}
\usepackage{cite}
\usepackage{authblk}

\begin{document}
\begin{titlepage}
\vspace*{2cm}
\begin{center}
\Large{\textbf{The collision rate ansatz for the classical Toda lattice}}\bigskip\\
{\large Herbert Spohn}
 \end{center}
 \begin{center}
Zentrum Mathematik and Physik Department, TUM,\\
Boltzmannstra{\ss}e 3, 85747 Garching, Germany,
 \texttt{spohn@tum.de}\\
 Department of Physics, Tokyo Institute of Technology
 \end{center}
\vspace{3cm}
\textbf{Abstract}: Considered is a generalized Gibbs ensemble of the classical Toda lattice.
We establish that the collision rate ansatz follows from (i)  the charge-current susceptibility matrix is symmetric
and (ii) the stretch current is proportional to the momentum, hence conserved. 
\end{titlepage}

\textit{Introduction}. -- While the idea of devising a hydrodynamic type theory for integrable many-body systems has been floating for a while \cite{Z71,E05},
the decisive breakthrough was accomplished only in 2016 \cite{CDY16,PNCBF17}. 
Integrable systems have an extensive number of conserved fields (= charges). Thus the notion of thermal equilibrium has to be
advanced to generalized Gibbs ensembles (GGE), which depend on an extensive number of chemical potentials. Conventional hydrodynamics is based on the propagation of local equilibrium states. Correspondingly, generalized hydrodynamics (GHD) relies on the propagation of local GGE states.  To obtain a coupled set of hyperbolic conservation laws, one has to first compute the GGE average of the charges. The crucial next step is to figure out the GGE averaged charge currents as a functional of the GGE averaged charges. Only then GHD is a closed set of dynamical equations.
In  \cite{CDY16,PNCBF17} a particular form of this functional has been proposed and we illustrate the argument directly for the 
classical Toda lattice, a chain with a purely exponential interaction potential.

\begin{figure}[!ht]
\centering
\includegraphics[width=0.75\columnwidth]{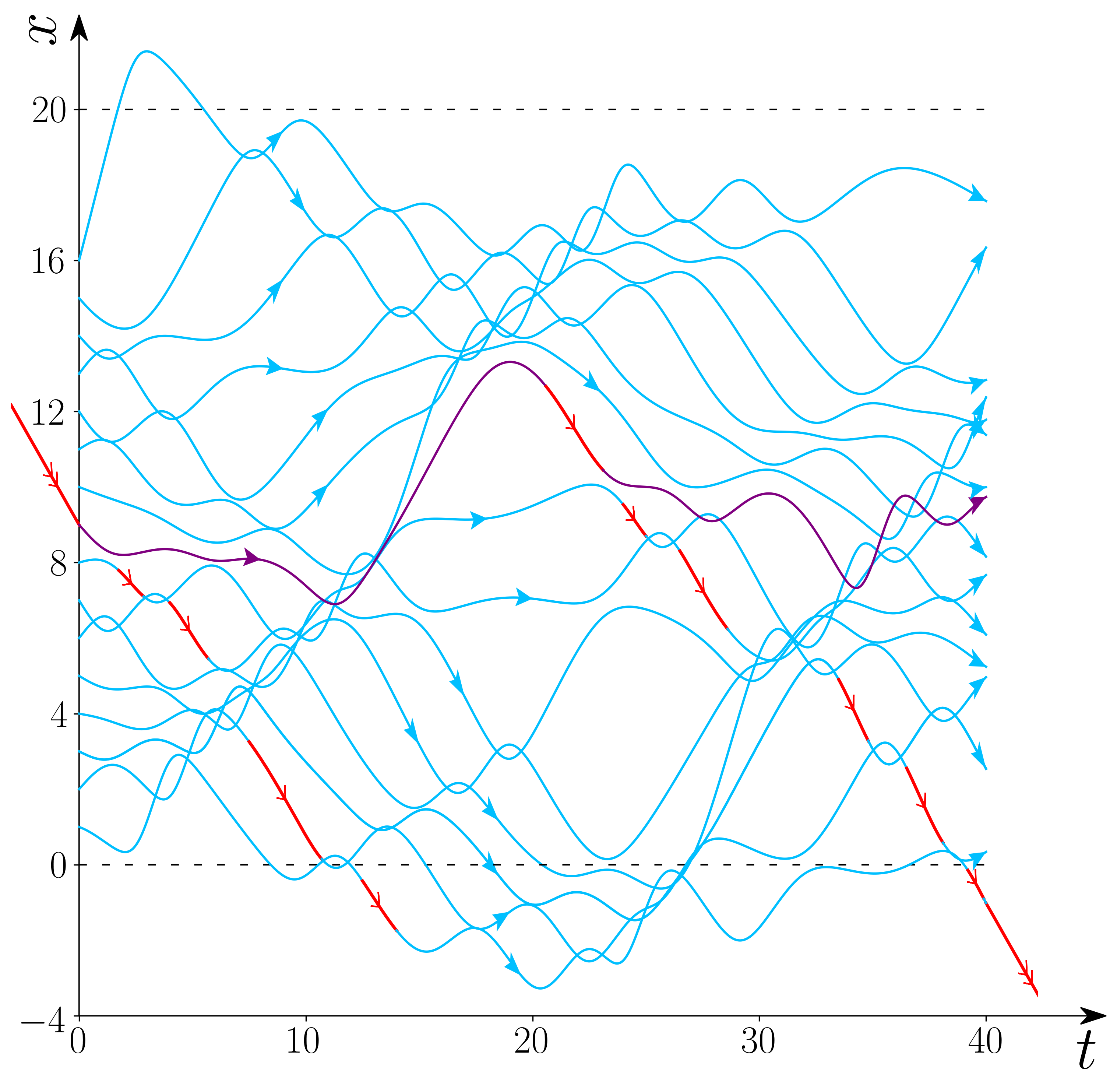}
\caption{In blue the trajectories of 16 Toda particles in a periodic box at density 0.8. One choice of the quasiparticle trajectory starting at 
the position of the 9-th particle
is shown in red. For comparison the trajectory of the 9-th particle is displayed in purple.}
\label{quasiparticle}
\end{figure}

To start, we assume that particles are far apart and the dynamics consists of essentially isolated two-particle collisions.
For the Toda lattice the two-particle phase shift is given by $\phi(v,v') = 2 \log|v - v'|$ referring  to a pair of incoming velocities $v,v'$. 
A quasiparticle coincides with a real particle for trajectory segments which have an essentially constant velocity, see Fig. \ref{quasiparticle}
lower right for a few two-particle collisions.
Thus a quasiparticle has some given velocity except for jumps either to the right or left during an encounter with another particle,
which can also be viewed as delay times of either sign. 
We now fix a GGE with some velocity density of quasiparticles, $\rho_\mathrm{p}(v)$. The average charge current is linked to the effective velocity of 
a single quasiparticle, which leads to the ansatz
\begin{equation}\label{1}
v^\mathrm{eff}(v) 
 = v + 2\int_\mathbb{R} \mathrm{d}v' \log|v-v'| \rho_\mathrm{p}(v')\big(v^\mathrm{eff}(v') - v^\mathrm{eff}(v)\big). 
\end{equation}
Here the bare velocity of the quasiparticle is denoted by $v$ which, because of collisions,  is modified to an effective velocity $v^\mathrm{eff}(v) $ on a mesoscopic scale. The integral sums over all collision partners: in a collision with a quasiparticle of velocity $v'$,
the first factor of the integrand is the jump size and the second factor the number of encounters per unit time. 

For the Toda lattice Eq. \eqref{1} seems to be  a reasonable approximation in case of low density. But the authors of  \cite{CDY16,PNCBF17}
assert the much stronger claim that the ansatz \eqref{1} is valid at \textit{any} particle density, even in the regime of strong interactions. 
An illustration is provided
in Fig. \ref{quasiparticle}, which displays the motion of 16 Toda particles in a periodic box at density $0.8$ in  units of the 
decay length of the exponential interaction. The segments of the quasiparticle trajectory are colored in red. One notes
well isolated two-particle encounters, but there are also more intricate multi-particle collisions with considerable 
arbitrariness in the precise choice of the quasiparticle trajectory. In \cite{CBS19} the ansatz has been checked numerically for a set of 
four distinct GGE parameters over the entire range of velocities. The agreement with \eqref{1} is extremely convincing and there is little doubt
that for the classical Toda lattice the ansatz is valid with  no exceptions. Still, from a theoretical perspective wanted would be an analytical argument, which is the goal of this letter. 

For a system of hard rods the analogue of the ansatz has been stated in \cite{P69},  being proved much later \cite{BDS83}. As major simplification, 
the hard rod two-particle phase shift is independent of the incoming velocities. Also the collision time is zero and thus the quasiparticle picture is exact. Related results have been recently obtained for the box-ball
model \cite{F18,CS20}, which is a space-time discrete dynamical system and has some similarity to a mixture of hard rods with arbitrary integer rod length. On the quantum side substantial  progress is reported for the spin-$\tfrac{1}{2}$ XXZ   chain \cite{VY19,BPP19,P19}. For more details I refer to the recent lecture notes  \cite{D19a}, which present a well accessible introduction to GHD, including a long list of references on the \medskip topic.

\textit{Conserved charges and their currents}. -- 
The Toda lattice is an anharmonic chain, for which the  interaction potential is specified to be exponential. The corresponding  hamiltonian is written as 
\begin{equation}\label{2}
H = \sum_{j\in\mathbb{Z}}\big( \tfrac{1}{2}p_j^2 + \mathrm{e}^{-r_j}\big),\quad r_j = q_{j+1} - q_j.
\end{equation}
Here $q_j,p_j$ is  position and momentum of the $j$-th particle.  The positional increments, $r_j$, are called stretches. We consider the 
infinitely extended lattice and indicate suitably finite lattice approximations whenever required. Then the equations of motion read
 \begin{equation}\label{3}
\frac{d}{dt}q_j = p_j, \qquad \frac{d}{dt}p_j =  \mathrm{e}^{-r_{j-1}}  -\mathrm{e}^{-r_j},\quad j \in \mathbb{Z},
\end{equation}
which are viewed as a discrete nonlinear wave equation.
Let us introduce the Flaschka  variables \cite{F74}, 
\begin{equation}\label{4}
a_j = \mathrm{e}^{-r_j/2},\quad b_j =  p_j.
\end{equation} 
In fact  the original Flaschka variables carry both a prefactor $\tfrac{1}{2}$ and, in principle, any common prefactor could be used. 
However, only with our choice the generalized free energy has a particularly simple form. The Lax matrix  is  the 
tridiagonal real symmetric matrix with matrix elements
\begin{equation}\label{5}
L_{j,j} = b_j, \quad L_{j,j+1} = L_{j+1,j}= a_j, 
\end{equation} 
and  $L_{i,j} = 0$ otherwise. For the finite lattice $[1,....,N]$, the $N$ eigenvalues of $L_N$ are conserved under the Toda time evolution. However, as functions on phase space they are non-local \cite{H74}. 
The approriate local version of the $n$-th charge is given by
$\mathrm{tr}[(L_N)^n]$, $n =1,2,...$,  which clearly has the density 
\begin{equation}\label{6} 
 Q^{[n]}_j = (L^n)_{j,j},\quad j \in \mathbb{Z}.
\end{equation}
By expanding the matrix element $(L^n)_{j,j}$, the density $Q^{[n]}_j$ depends at most on the variables $\{a_i,b_i, i \in [n-j,...,n+j]\}$, hence is strictly local.  
In addition, the stretch is conserved and has the density
\begin{equation}\label{7} 
 Q^{[0]}_j = r_j. 
\end{equation}
From the equations of motion, the lattice currents are defined through
\begin{equation}\label{8} 
 \tfrac{d}{dt} Q^{[n]}_j = J^{[n]}_{j} - J^{[n]}_{j+1} 
\end{equation}
yielding the respective current densities
\begin{equation}\label{9} 
J^{[0]}_j = - Q^{[1]}_j , \qquad J^{[n]}_j = \tfrac{1}{2}(L^nL^\mathrm{off})_{j,j},
\end{equation}
where $L^\mathrm{off}$ denotes the off-diagonal part of $L$ \cite{S19}\medskip.

\textit{Generalized hydrodynamics}. -- 
We will need a few standard items from generalized hydrodynamics. A generalized Gibbs 
state (GGE) of the Toda lattice is characterized by the pressure $P > 0$
and the chemical potential $V(w)$, which one should think of as a finite polynomial with $\lim_{|w| \to \infty} V(w) = \infty$.
For the finite volume $[1,...,N]$ and free boundary conditions, $(L_N)_{1,N} = 0= (L_N)_{N,1}$,  the GGE in the variables of \eqref{4} is defined through
the probability density
\begin{equation}\label{10}
\frac{1}{Z_\mathrm{toda}(N,P,V)}\prod_{j=1}^{N}  \mathrm{d}b_j  \prod_{j=1}^{N-1} \mathrm{d}a_j \frac{2}{a_j}
(a_j)^{2P} \mathrm{e}^{-\mathrm{tr}[V(L_N)]}, \quad b_j \in \mathbb{R}, \,\,a_j \in \mathbb{R}_+,
\end{equation}
with the normalizing partition function $Z_\mathrm{toda}(N,P,V)$. The Boltzmann weight is the exponential of 
\begin{equation}\label{11}
\mathrm{tr}[V(L_N)] = \sum_{n=1}^{\bar{n}} \mu_n \mathrm{tr}[(L_N)^n].
\end{equation}
Following the rules of statistical mechanics, the $n$-th charge is controlled by the chemical potential $\mu_n$. The pressure $P$ is the thermodynamic dual of the stretch.
Expectations with respect to \eqref{10} are denoted by 
$\langle \cdot \rangle_{N,P,V}$. The infinite volume measure exists, expectations being denoted by  $\langle \cdot \rangle_{P,V}$,
and the limit measure has correlations decaying exponentially, see  \cite{S19} and Appendix of \cite{S19a}.

The GGE average stretch is given by 
\begin{equation}\label{12}
\lim_{N\to \infty} \frac{1}{N} \sum_{j=1}^N\langle r_j \rangle_{N,P,V} = \langle  r_0 \rangle_{P,V} =  \nu
\end{equation}
and the GGE average of the $n$-th charge by
\begin{equation}\label{13}
\lim_{N\to \infty} \frac{1}{N} \langle \mathrm{tr}[(L_N)^n] \rangle_{N,P,V} = \langle  (L^n)_{0,0} \rangle_{P,V} =  \int_\mathbb{R}  \mathrm{d}w \rho_\mathrm{Q}(w) w^n,
\end{equation}
which defines $\rho_\mathrm{Q}$. Clearly $\rho_\mathrm{Q}(w)$ is the density of states (DOS) of the Lax matrix. By construction  $\rho_\mathrm{Q}(w) \geq 0$ and $\int_\mathbb{R} \mathrm{d}w \rho_\mathrm{Q}(w) =1$. 
Correspondingly we introduce the GGE average of the $n$-th current  by
\begin{equation}\label{14}
\lim_{N\to \infty} \frac{1}{N}\tfrac{1}{2} \langle \mathrm{tr}[(L_N)^nL_N^\mathrm{off}] \rangle_{N,P,V}   = \tfrac{1}{2}\langle  (L^nL^\mathrm{off})_{0,0} \rangle_{P,V} = \int_\mathbb{R} \mathrm{d}w \rho_\mathrm{J}(w)w^n,
\end{equation}
which defines the current DOS  $\rho_\mathrm{J}$.  Setting $n=0$, one concludes that $\int_\mathbb{R}  \mathrm{d}w \rho_\mathrm{J} (w)
 =0$. Hence the current DOS has no definite sign\medskip.\\
\textit{Note}: We use standard notations from GHD. Only  the conventional state density, $n$, collides with the index $n$ and $\rho_\mu$ is taken instead.\medskip

Let us define the integral operator 
\begin{equation}\label{15}
T\psi(w) = 2 \int_\mathbb{R} \mathrm{d}w' \log |w-w'| \psi(w'),\quad w \in \mathbb{R}.
\end{equation}
Then  $\rho_\mu$ is the solution of the classical version of the TBA equation,
\begin{equation}\label{16} 
  V(w) -  T\rho_\mu(w) +\log \rho_\mu(w)  - \mu = 0,
 \end{equation}
with $\mu$ adjusted such that $\int_\mathbb{R} \mathrm{d}w \rho_\mu(w) =P$. The dressing of a function $\psi$  is defined by
\begin{equation}\label{17} 
\psi^\mathrm{dr} = \psi + T \rho_\mu \psi^\mathrm{dr},\quad \psi^\mathrm{dr} = \big(1 - T\rho_\mu\big)^{-1} \psi.
\end{equation}
On the right hand side $\rho_\mu$ is regarded as multiplication operator, $(\rho_\mu\psi)(w) = \rho_\mu(w)\psi(w) $.
Differentiating TBA with respect to $\mu$ one concludes
\begin{equation}\label{18} 
\rho_\mathrm{p}= (1 - \rho_\mu T)^{-1} \rho_\mu = \rho_\mu(1 - T\rho_\mu)^{-1}[1] = \rho_\mu[1]^\mathrm{dr}.
 \end{equation}
 Here $[1]$ stands for the constant function, $\psi(w) = 1$, and similarly $[w^n]$ for the $n$-th power, $\psi(w) = w^n$.
 
 As shown in \cite{BCM19,D19,S19}, the charge DOS is given by
 \begin{equation}\label{19} 
\rho_\mathrm{Q}(w) = \nu \rho_\mathrm{p}(w).
 \end{equation}
 A particularly simple example is thermal equilibrium, which corresponds to
 $V(w) = \tfrac{1}{2} \beta w^2$. Then the Lax matrix has statistically independent matrix elements:  The diagonal $b_j$'s are mean zero Gaussian  with variance $\beta^{-1}$ and the off-diagonal $a_j$'s are $\chi$-distributed with parameter $P$, i.e. for the even moments $\langle (a_j)^{2n}\rangle = P(P+1)...(P+n-1)$. In this case an analytical formula for $\nu \rho_\mathrm{p}$ is available 
 \cite{Opper,ABG12}. But for general GGE one has to rely on numerical simulations, either by directly sampling the Lax matrix or by solving TBA, see \cite{CBS19} for a few examples.
 
 Assuming the collision rate ansatz, the charge DOS is written as 
 \begin{equation}\label{20} 
\rho_\mathrm{J}(v) =  \rho_\mathrm{p}(v)( v^\mathrm{eff}(v) -q_1)
 \end{equation}
 with $v^\mathrm{eff}(v)$ the solution to \eqref{1} and the average momentum $q_1 = \nu \int_\mathbb{R} \mathrm{d}w \rho_\mathrm{p}(w)w$ \cite{D19}.
 As observed in \cite{CDY16,PNCBF17}, see  also Appendix of \cite{CBS19} for the Toda lattice,  $v^\mathrm{eff}(v)$ can be rewritten more
 concisely as $ v^\mathrm{eff} = [v]^\mathrm{dr}/[1]^\mathrm{dr}$.
 Thus the collision rate ansatz states that
  \begin{equation}\label{21} 
 \rho_\mathrm{J}(v) = \rho_\mu (v)[v]^\mathrm{dr}(v) - q_1 \rho_\mathrm{p}(v)
 \end{equation}
 and it is this identity which has to be established on the basis of the microscopic definition \eqref{14}.
 
Our discussion suggest the following physical picture. In the initial state the DOS of the Lax matrix and the local stretch are assumed 
to vary slowly on the microscopic scale and hence depend on the hydrodynamic spatial scale. Because of the conservation laws
both quantities move then also slowly in time. Eq. \eqref{21} provides the local currents as a functional of the local DOS and stretch.
Solving GHD with these currents then tells us how the local charges change on the hydrodynamic spacetime scale.

Since in thermal equilibrium the random Lax matrix has such a simple structure, one might be tempted to fully expand the term
$(L^n L^\mathrm{off})_{0,0}$ as an $n$-step random walk and to insert the path sum on the left of Eq. \eqref{14}.  For the right hand side 
of \eqref{21} one could use the explicit expression for $\rho_\mathrm{p}$.
This works well for small $n$, say up to $n=5$, and 
confirms the ansatz. But for larger $n$ the combinatorics becomes intricate and it seems difficult to identify any recursive  \medskip patterns.

\textit{Proof of the collision rate ansatz}. --
We first introduce a convenient shorthand. Consider extensive observables, say $Q,\tilde{Q}$, with strictly local
 densities  $Q_j,\tilde{Q}_j$, i.e. formally $Q = \sum_{j \in \mathbb{Z}} Q_j$ and the same for $\tilde{Q}$. Then the respective susceptibility is denoted by  
\begin{equation}\label{22}
\big\langle Q;\tilde{Q}\big\rangle_{P,V} = \sum_{j \in \mathbb{Z}} \big(\big\langle Q_0 \tilde{Q}_j\big\rangle_{P,V} - \big\langle Q_0\big \rangle  \big\langle \tilde{Q}_0\big\rangle_{P,V}\big).
\end{equation}
We also introduce the charge-current susceptibility matrix through
\begin{equation}\label{23}
B_{m,n} = \big\langle Q^{[m]};J^{[n]}\big\rangle_{P,V}
\end{equation}
and note the symmetry
\begin{equation}\label{24}
B_{m,n} = B_{n,m}, \quad m, n= 0,1,...\,.
\end{equation}
This symmetry is a very general property of local conservation laws.  Required is only the invariance of the reference state $\langle \cdot \rangle$ under space-time translations together with a clustering of the static two-point function appearing on the right side of  \eqref{22}
\cite{DBD19}.

Now let us consider the $P$-derivative of the average current 
 \begin{equation}\label{25}
   \partial_P \tfrac{1}{2}\langle( L^nL^\mathrm{off})_{0,0} \rangle_{P,V} = - B_{0,n} = - B_{n,0}
   = \partial_{\mu_1} \langle Q^{[n]}\rangle_{P,V},
\end{equation}
according to \eqref{24} and  since $J^{[0]} = - Q^{[1]}$.  We integrate both sides of \eqref{25} in $P$ and note that the average current vanishes at $P=0$. 
  Then, as discussed in \cite{S19}, the $P$-integral 
on the right hand side accounts for the 
ramping when mapping to the Dumitriu-Edelman random matrix \cite{DE02}.  Thus the limit of the right hand side translates to the corresponding fluctuation covariance of the 1D log gas with a confining potential $V$ and interaction strength $P/N$, i.e. in the mean field regime. This problem is discussed in Section 5 of \cite{S19a} and we merely have to copy Eq. (5.11) from there with the result
\begin{eqnarray}\label{28}
&&\hspace{-40pt} \tfrac{1}{2}\langle( L^nL^\mathrm{off})_{0,0}\rangle_{P,V} = \lim_{N\to \infty} \frac{1}{N} \int _0^P \hspace*{-4pt}\mathrm{d} u 
 \langle \mathrm{tr}[L_N - q_11\hspace*{-3pt}\mathrm{l}_N] \mathrm{tr}[(L_N)^n]\rangle_{N,u,V}\nonumber\\
  &&\hspace{40pt}= \int_\mathbb{R} \mathrm{d} w w\big((1- \rho_\mu T)^{-1} \rho_\mu
 [w^n]\big)(w) - q_1 \int_\mathbb{R} \mathrm{d} w\rho_\mathrm{p}(w)w^n, 
\end{eqnarray}
which is the required identity \eqref{21} by noting that $(1- \rho_\mu T)^{-1} \rho_\mu$ is a symmetric \medskip operator. 
 
\textit{Discussion}. -- Our result is surprising, since only the generic symmetry \eqref{24} comes into action,
except for some identities related to the generalized free energy of the Toda chain. Stepping back from the Toda lattice, and considering the class of integrable many-body systems, the cornerstone of our argument is the availability of a self-conserved current, i.e. a current which itself is a locally conserved charge. In this form the generalization to quantum systems looks doable. Indeed, the symmetry of the 
charge-current susceptibility matrix holds in generality.
For the  Lieb-Liniger $\delta$-Bose gas the particle current  is self-conserved, since it equals the total momentum.  As well known, for XXZ spin chain the energy current is self-conserved. With suitable adaptions the collision rate ansatz can be directly verified for both models \cite{YS20}. On the other hand, the integrable spin-$\tfrac{1}{2}$ Fermi-Hubbard model has no self-conserved charge \cite{ID17}.  Thus it would be of interest to understand how our strategy extends to further integrable many-body systems.

The reader might have noticed that our notation switches from the $v$-argument to the $w$-argument. This is not accidental. 
 In the collision rate ansatz \eqref{1} we argued on the basis of  low density and $v$ refers to the velocity of quasiparticles. But in our analytical argument only the spectral density of the Lax matrix appears. Hence $w$ is interpreted as the parameter of the spectral density, i.e. the Lax DOS.
  Our
 argument can also be read in reverse order. Starting from \eqref{25}, the formula for the GGE averaged currents is derived, which in retrospect is then reshuffled  into the collision rate ansatz.\\\newpage\noindent
 \textbf{Acknowledgements}. This article was written during my stay at the Tokyo Institute of Technology in the spring 2020.
 I am most grateful to Tomohiro Sasamoto for his warm hospitality. I thank Atharv Deokule for producing the figure and also 
 Sylvain Prolhac and Tomohiro Sasamoto for their insights on the combinatorics of weighted random walks arising from the 
 random Lax matrix connected to thermal equilibrium.\\\\


\begin{thebibliography}{99}
 
\bibitem{Z71}V. E. Zakharov, Kinetic equation for solitons, Sov. Phys. JETP \textbf{33}, 538 (1971).

\bibitem{E05} G. A. El and A. M. Kamchatnov, Kinetic equation for a dense soliton gas, Phys. Rev.
Lett. \textbf{95}, 204101 (2005).

\bibitem{CDY16} O. A. Castro-Alvaredo, B. Doyon, and T. Yoshimura, Emergent hydrodynamics in integrable quantum systems out of equilibrium, Phys. Rev. X \textbf{6}, 041065 (2016).

\bibitem{PNCBF17} L. Piroli, J. De Nardis, M. Collura, B. Bertini, and M. Fagotti, Transport in out-of-equilibrium XXZ chains: Nonballistic behavior and correlation functions, Phys. Rev. B \textbf{96}, 115124 (2017).

\bibitem{CBS19}  Xiangyu  Cao, V. Bulchandani, and H. Spohn, The GGE averaged currents of the classical Toda chain, arXiv:1905.04548.

\bibitem{P69} J. K. Percus, Exact solutions of kinetics of a model classical fluid, Phys. Fluids \textbf{8}, 1560 (1969).    

\bibitem{BDS83} C. Boldrighini, R. L. Dobrushin and Yu. M. Sukhov,
	One-dimensional hard rod caricature of hydrodynamics,
	J. Stat. Phys. {\bf 31}, 577 (1983).
	
\bibitem{F18} P. A. Ferrari, C. Nguyen, L. Rolla, and M. Wang, Soliton decomposition of the box-ball
system, arXiv:1806.02798.	

\bibitem{CS20} D. Croydon, M. Sasada, Generalized hydrodynamic limit for the box-ball system,
arXiv:2003.06526.  

\bibitem{VY19} D.-L. Vu, T. Yoshimura, Equations of state in generalized hydrodynamics, arXiv:1809.03197.

\bibitem{BPP19} M. Borsi, B. Pozsgay, and L. Pristy\'{a}k, Current operators in Bethe ansatz and generalized hydrodynamics: an exact quantum/classical correspondence, arXiv:1908.07320.

\bibitem{P19} B. Pozsgay, Current operators in integrable spin chains: lessons from long range 
deformations, arXiv:1910.12833.

\bibitem{D19a} B. Doyon, Lecture Notes on Generalised Hydrodynamics, arXiv:1912.08496.

\bibitem{F74} H. Flaschka, The Toda lattice. II. Existence of integrals, Phys. Rev. B \textbf{9}, 1924 (1974).

\bibitem{H74} M. Henon, Integrals of the Toda lattice, Phys. Rev. B \textbf{9}, 1921 (1974).

\bibitem{S19} H. Spohn, Generalized Gibbs ensembles of the classical Toda chain,\\ arXiv:1902.07751v3,  to appear J. Stat. Phys. (2020).

\bibitem{S19a} H. Spohn, Ballistic space-time correlators of the classical Toda lattice,\\ arXiv:1911.10825.

\bibitem{BCM19} V. Bulchandani, Xiangyu  Cao, and J. Moore, Kinetic theory of quantum and classical Toda lattices, 
J. Phys. A \textbf{52}, 33LT01 (2019).

\bibitem{D19} B. Doyon, Generalised hydrodynamics of the classical Toda system, J. Math. Phys. \textbf{60}, 073302 (2019).

\bibitem{Opper} M. Opper, Analytical solution of the classical Bethe-ansatz solution for the Toda chain, Phys. Lett. A \textbf{112}, 201 (1985).

\bibitem{ABG12} R. Allez, J.-P. Bouchaud, and A. Guionnet, Invariant $\beta$-ensembles and the Gauss-Wigner crossover,
Phys. Rev. Lett. \textbf{109}, 094102 (2012).

\bibitem{DE02} I. Dumitriu and A. Edelman, Matrix models for beta ensembles, J. Math. Phys. \textbf{43}, 5830 (2002).

\bibitem{DBD19} J. De Nardis, D. Bernard, and B. Doyon, Diffusion in generalized hydrodynamics and quasiparticle scattering,
SciPost Phys. \textbf{6}, 049 (2019).

\bibitem{YS20} T. Yoshimura and H. Spohn, Collision rate ansatz for quantum integrable systems, in preparation (2020).

\bibitem{ID17} E. Ilievski and J. De Nardis, Ballistic transport in the one-dimensional Hubbard model: the hydrodynamic approach, Phys. Rev. B {\bf 96}, 081118 (2017).



\end{thebibliography}
\end{document}